\documentclass[twocolumn,showpacs,preprintnumbers,amsmath,amssymb]{revtex4}
\usepackage{graphicx}
\usepackage{color}
\def\beq{\begin{equation}}
\def\eeq{\end{equation}}
\def\beqa{\begin{eqnarray}}
\def\eeqa{\end{eqnarray}}
\def\bfx{\mbox{\boldmath $x$}}

\def\bfv{\mbox{\boldmath $v$}}
\def\bfJ{\mbox{\boldmath $J$}}

\begin{document}
\title{Oscillation modes, Decoherence, Beam Transfer Functions, and Instabilities for a Beam Subjected to 
Space-Charge Forces}
%
\author
{
Kazuhito Ohmi, Liangsheng Huang, 
Hanyang Liu, Li Rao, Renjun Yang, Sheng Wang and Mingyang Huang\\
\it{Institute  of High Energy Physics, Chinese Academy of Sciences, 
Beijing 100049, China}\\
CSNS,  Spallation Neutron Source Science Center, Dongguan 523803, China
}
\begin{abstract}  
Based on the oscillation modes of a bunch subject to the space-charge force,
we discuss the decoherence of the dipole oscillation, the beam transfer function,
and the beam instability.
The longitudinal phase space is discretized into \(n_l\times n_J\) grids in the azimuthal and
radial directions, respectively.
The transverse dipole moment in the longitudinal phase space
is represented by a vector of length \(n_l n_J\), and
the transformation given by the Vlasov equation is represented by a matrix transformation
of size \(n_l n_J\times n_l n_J\).
Solving the eigenvalue problem of this matrix gives the \(n_l n_J\) eigen-oscillation modes of the bunch.
Using the resulting eigenvectors within a standard linear-algebraic treatment,
we analyze these three phenomena.
The analysis was performed for the Rapid Cycling Synchrotron (RCS) of the China Spallation Neutron Source (CSNS)
using \(n_l=41\) and \(n_J=40\),
and it visualizes the features related to the continuous distribution of the tune,
Landau damping, and isolated modes separated from the continuous tune region.
\end{abstract}
\maketitle

\section{Introduction}
The behavior of beam particles moving in a circular accelerator can be described by 
equations of motion and/or a Hamiltonian. 
The primary objective is to determine $\bfx(s)$ at a given $s$ for an initial set of 
coordinates and momenta $\bfx(s=0)$. 
We generally deal with the six-dimensional phase space represented by $(x,p_x,y,p_y,z,\delta)$. 
Here, to handle coupled oscillations in the transverse $x$ and longitudinal $z$ directions, 
we use the four-dimensional subspace $(x,p_x,z,\delta)$. 
The $y-z$ oscillation can be treated similarly to the $x-z$ oscillation.

To investigate the collective behavior of a beam, we consider the phase-space distribution of 
particles, $\Psi(\bfx)$. 
The motion of a single particle is described by equations of motion and a Hamiltonian, 
$H(\bfx)$, which depend only on the phase-space coordinates of that particle; 
the effects of the other particles are incorporated by averaging. The equation of motion governing 
$\Psi(\bfx)$ is called the Vlasov equation and is expressed as follows:
\beq
\frac{\partial \Psi}{\partial s}+
\frac{\partial \Psi}{\partial x}\frac{\partial H}{\partial p_x}
-\frac{\partial \Psi}{\partial p_x}\frac{\partial H}{\partial x}
+
\frac{\partial \Psi}{\partial z}\frac{\partial H}{\partial \delta}
-\frac{\partial \Psi}{\partial \delta}\frac{\partial H}{\partial z}
=0. \label{VlasovEq}
\eeq
The Vlasov equation can also be expressed using Poisson brackets:
\beq
\frac{\partial \Psi}{\partial s}=[H,\Psi]=:H:\Psi. \label{VEpoi}
\eeq
Here, the Poisson bracket is expressed by the operator $::$.
A formal solution can be written as:
\beq
\Psi(\bfx;s)=\exp(:H:s)\Psi(\bfx;0).\label{VEsol}
\eeq
Equation (\ref{VEpoi}) has the opposite sign of $H$ compared to Hamiltonian equations of 
motion:
\beq
\frac{d \bfx}{d s}=-[H,\bfx]. \label{VEpoi}
\eeq
This is a difference in perspective: focusing on the particle
distribution versus on the motion of individual particles.
The distribution of a single particle can be represented using the
Dirac delta function as
\beq
\Psi(\bfx;0)=\delta(\bfx-\bfx_i),
\eeq
and tracking the distribution using the Vlasov equation yields
\beq
\Psi(s)=\delta(e^{:H(\bfx):s}\bfx-\bfx_i)=\delta(\bfx-e^{-:H(\bfx_i):s}\bfx_i).
\eeq

It can be seen that the transformation of the individual particle
coordinates $\mathbf{x}_i$ follows Eq.~(\ref{VEpoi}), meaning that the
distribution changes due to particle motion.
In other words, the Vlasov equation states that the current phase-space
density equals the past phase-space density, as the past density moves
along with the particle motion.
In Hamiltonian dynamics, the simplecticity ensures that the Jacobian of 
the transformation $e^{:H:}$ is unity.
The premise is that the phase-space density does not change with motion.
The Vlasov equation is derived from a single-particle Hamiltonian for
independent motion within an averaged field, and short-range two-particle
collisions such as intrabeam scattering are not considered.

Eq.~(\ref{VEsol}) is represented by a matrix transformation once the phase space is discretized.
However, if one attempts to discretize the six-dimensional phase space, the matrix becomes
extremely large. Therefore, by restricting the treatment to the dipole mode \(x(z,\delta)\)
in the \(x\) direction alone, it suffices to discretize the longitudinal phase space.
Here, for a beam subject to the space-charge force, we obtain the eigen-oscillations and
eigenvectors (eigen-oscillation distributions) by solving the eigenvalue problem for the
transformation matrix of the dipole-mode distribution, represented by a vector obtained by
discretizing the longitudinal phase space into \(41\ (\text{azimuthal})\times40\ (\text{radial})=1640\),
i.e., a matrix of size \(1640\times1640\).

Starting from the oscillation modes of a bunch undergoing
synchro-betatron oscillations in an accelerator, we discuss the oscillation
modes of a bunch oscillating under the influence of space-charge forces.
As application examples, we discuss (i) the oscillations initialized by a
uniform offset, (ii) the oscillations under sinusoidal
forced excitation, i.e., the beam transfer function, and (iii) 
the beam istability that appear when an impedance is added.

Studies on the head-tail oscillation modes and instabilities have been carried out in 
many papers~\cite{MBlask1,burov2009,Kornilov2010,AMac,chin,XBuf}.
They were based on the airbag square-well (ABS) model in the longitudinal phase space,
which is easy to handle analytically.
Numerical studies extending this approach have been carried out in Ref.~\cite{AMac}.
Experiments and analyses for (i) kicker excitation and (ii) forced oscillation
have been carried out based on the ABS model in
Refs.~\cite{Yoshimura} and \cite{Kornilov2010,Singh2013}.
Coupled-bunch instabilities are analyzed based on the ABS model in Ref.~\cite{burovCBI}.
The numerical treatment using the circulant matrix model is presented in Ref.~\cite{XBuf}.

The analysis presented here has been used to study coupled-bunch instability~\cite{cbisc}.
Here we focus on the \(1640\) oscillation modes (eigenvectors) obtained
by solving the eigenvalue problem of a \(1640\times 1640\) matrix.
Our method allows us to handle arbitrary longitudinal distributions.
The above subjects (i)-(iii) are discussed by a unified linear-algebraic technique;
Landau damping is visualized in all three subjects.

This paper is organized as follows.
The particle distribution for free synchro-betatron motion is analyzed in
Sec.~\ref{sbosc}: the basic formalism is presented in Sec.~\ref{sbosca},
and the application examples (i) and (ii) are discussed in
Secs.~\ref{sboscb} and \ref{NoSCBTF}, respectively.
The particle distribution with space-charge forces is analyzed in
Sec.~\ref{scosc}: the basic formalism is presented in Sec.~\ref{scosca} and Sec.~\ref{scoscb},
and the application examples (i), (ii), and (iii) are discussed in
Secs.~\ref{scoscc}, \ref{SCBTF}, and \ref{scinstab}, respectively.

For a numerical analysis, the parameters of the China Spallation Neutron
Source (CSNS) are used, as listed in Table~\ref{accpara}.

\begin{table}
\begin{tabular}{l|c|cccc}
\hline
Circumference & $C$ (m) & 227.9 \\
Injection Energy & $E_0$ (GeV)  & 0.08 \\
Bunch population & $N_p$ &  $0.75\times 10^{13}$ \\
Bunch length & $\sigma_z$ (m) & 25  \\
Slippage factor & $\eta_p$ & $-0.81$ \\
Synchrotron tune & $\nu_s$ &  0.01 \\
Chromatic phase & $\chi/\xi$ & $-1.2$ \\
Space charge tune shift & $\Delta\nu_{sc}$ & $-0.27$ \\
\hline
\end{tabular}
\caption{Parameters of CSNS used as a numerical application. }\label{accpara}
\end{table}

\section{Synchro-Betatron Oscillations}\label{sbosc}
\subsection{Distribution of particles undergoing free synchro-betatron oscillations}\label{sbosca}
The solution in the linear dynamics of a circular accelerator is well known.
There are invariants in each degree of freedom, as
\beq
J_x=\frac{\gamma_x x^2+2\alpha_x xp_x+\beta_x p_x^2}{2},
\quad J=\frac{z^2/\beta_z+\beta_z\delta^2}{2}
\eeq
For \(z\), we do not consider the tilt of the \(z\)-\(\delta\) phase space; $\alpha_z=0$.
$\beta_z$ is defined as the aspect ratio of the \(z\)-\(\delta\) ellipse;
$\beta_z=z_{\max}/\delta_{\max}$ in the phase-space trajectories.
Using the cavity parameters and the slippage factor (\(\eta_p\)), it is expressed as
$\beta_z=\sqrt{|\eta_p| C \beta^2E_0/(eV')}/(2\pi)$, independent of \(s\),
where \(C\) is the circumference of a circular accelerator.
For simplicity, we denote $J=J_z$ and $\phi=\phi_z$ for the synchrotron phase-space variables.

The Hamiltonian is expressed as
\beq
H(\bfJ;s)=\frac{J_x}{\beta_x(s)}-\frac{\eta_p J}{\beta_z}
\eeq
or, considering the change in one revolution,
\beq
\bar{H}(\bfJ)\equiv \oint ds H(\bfJ;s)=2\pi\nu_xJ_x+2\pi\nu_z J.
\eeq
The solution to the equation of motion shows that the betatron and synchrotron
phases change in one revolution as
$\Delta \phi_x=2\pi\nu_x=\oint ds/\beta_x(s)$ and
$\Delta \phi_z=2\pi\nu_z=\eta_p \beta_z C $.

The stationary solution is a distribution function $\Psi(\bfJ)$ that depends only on \(J\).
\[
\Psi(\bfJ,C)=\exp(:H_{rev}(\bfJ):)\Psi(\bfJ,0)=\Psi(\bfJ,0).
\]
The Poisson bracket of functions of the same variable vanishes, $:J:\psi(J)=0$.

The betatron variables are replaced by
\beq
\frac{x}{\sqrt{\beta_x}}\Rightarrow x\qquad
\frac{\beta_x p_x+\alpha_x x}{\sqrt{\beta_x}}\Rightarrow p_x.
\eeq
Thus, they are represented by $J_x=(x^2+p_x^2)/2$, $x=\sqrt{2J_x}\cos\phi_x$,
$p_x=-\sqrt{2J_x}\sin\phi_x$.
Furthermore, we introduce $v=x+ip_x$.
With this, betatron oscillation as a function of \(s\) is represented by
a simple phase rotation:
\beq
v(s)=e^{-i\phi_x}v(0),\quad v(C)=e^{-2\pi i\nu_x}v(0).
\eeq
While $\bfx(s)$ is used as a function of \(s\) to describe motion,
$\bfx$ is also used as coordinates in phase space.
Using the longitudinal coordinates \((z,\delta)\) or \((J,\phi)\) as
phase-space coordinates,
we treat the evolution of the averaged transverse phase-space variables,
\beqa
v(z,\delta;s)&=&\frac{\int (x+ip_x)\Psi(\bfx;s)\,dx\,dp_x}{\psi(z,\delta)} \nonumber \\
\rho_v(z;s)&=&\int (x+ip_x)\Psi(\bfx;s)\,dx\,dp_x\,d\delta \label{vaver}\\
\langle v(s)\rangle  &=&  \int \rho_v(z;s)\,dz,\qquad
\bar{v}(z;s)=\frac{\rho_v(z;s)}{\rho(z;s)}\nonumber
\eeqa
where $\psi(z,\delta;s)=\int \Psi(\bfx;s)\,dx\,dp_x$ and
$\rho(z;s)=\int \psi(z,\delta;s)\,d\delta$
are the longitudinal beam densities.
\(s\) is sometimes omitted when it is trivial.

Non-stationary oscillations encompass any distribution not expressible solely by $\bfJ$.
Consider the case where the initial transverse distribution is given by $f(J_x)$ in the \(x\)-direction,
with its center of mass in the longitudinal phase space given by
$v(J,\phi;0)=x_0(J,\phi)$, and the longitudinal distribution is $\psi(J)$
around the center.
\beq
\Psi(\bfx;0)=f(x-x_0(J,\phi),p_x)\psi(J)
\eeq
in the \(x\)-\(p_x\) and \(z\)-\(\delta\) spaces.
We can find $\int\Psi\,dx\,dp_x=\psi(J)$.
The distribution $\Psi$ is not an equilibrium distribution for $x_0\ne0$,
since it is not a function of $\bfJ$.
As a typical example, in the case of a Gaussian in both planes, $f_G$ and $\psi_G$ are
\beqa
f_G(x-x_0(J,\phi),p_x)&=&\frac{\exp
\left(\frac{-(x-x_0(J,\phi))^2-p_x^2}{2\varepsilon_x}\right)}{2\pi\varepsilon_x},\\
\psi_G(J)&=&\frac{e^{-J/\varepsilon_z}}{2\pi\varepsilon_z}.
\eeqa
Integration over \(x\) and \(p_x\), after redefining the variable
$x-x_0(J,\phi)\rightarrow x$, gives, as expected,
\beq
x(J,\phi;0)=\int dx\,dp_x\,x\, f(x-x_0(J,\phi),p_x)=x_0(J,\phi),
\eeq
where \(f\) is an even function of the arguments $x-x_0$ and $p_x$; namely
$\int dx\,dp_x\,x f(x,p_x)=0$.

After one turn, the distribution function undergoes the inverse transformation
$e^{:H:}$ on $\bfx$,
changing the betatron and synchrotron phases by $-2\pi\nu_x$ and $-2\pi\nu_z$:
\beq
\Psi(\bfx;C)=f(e^{2\pi i\nu_x}v-x_0(J,\phi-2\pi\nu_z))\psi(J).
\eeq
By redefining the variables,
\beq
\Psi(\bfx;C)=f(v-e^{-2\pi i\nu_x}x_0(J,\phi-2\pi \nu_z))\psi(J).
\eeq
Thus, the center-of-mass position in phase space after one revolution is
\beq
v(J,\phi;C)=\int v f\,dx\,dp_x=e^{-2\pi i\nu_x}x_0(J,\phi-2\pi \nu_z).
\eeq

Here we introduce the synchro-betatron oscillation mode.
Since \(v\) is periodic with respect to the synchrotron phase, the azimuthal
direction in phase space, we perform a Fourier expansion into azimuthal modes:
\beq
v(J,\phi;s)=\sum_{l=-\infty}^{\infty} v_l(J;s)e^{il\phi}.\label{lmode}
\eeq
Focusing on a mode with \(l\),
\[
v(J,\phi;C)= e^{-2\pi i(\nu_x+l\nu_z)}x_0(J,\phi).
\]
The same argument holds if the initial offset is taken to be $v_0(J,\phi)$
instead of $x_0(J,\phi)$.
Eq.~(\ref{lmode}) means that the azimuthal dependence of the distribution,
$v\sim e^{il\phi}$ in the longitudinal phase space, is an eigenmode
of the free synchro-betatron motion, with eigenfrequency (tune) $\nu_x+l\nu_z$.


\subsubsection{Chromaticity}\label{Chromsec}
Chromaticity is a quantity characterizing the dependence of the betatron tune
on the momentum deviation (\(\delta=\Delta p/p_0\)) in single-particle motion.
The tune is expanded with respect to the momentum deviation, and its coefficient
is called the chromaticity.
Here, only the first-order term is considered and is denoted by the linear
chromaticity \(\xi\):
\beq
\Delta \nu_x(\delta)=\xi \delta.
\eeq

In the case of a finite chromaticity, the \(l\)-mode in the expansion of
Eq.~(\ref{lmode}) is no longer an eigenmode.
The \(l\)-mode including the variation due to \(\chi\)
becomes an eigenmode of oscillation, with tune \(\nu_x+l\nu_z\)~\cite{cbisc}:
\beq
v(z,\delta)= \sum_l v_l(J) e^{il\phi+i\chi z/\sigma_z}.\label{chrmlmode}
\eeq
where the chromatic phase \(\chi=2\pi\xi\sigma_z/(\eta_p C)\) is introduced.
The variation due to chromaticity within the bunch
is characterized by \(\chi\).
When the energy is below the transition, \(\eta_p\) is negative,
and the sign of \(\chi\) is opposite to the sign of the chromaticity \(\xi\).


\subsection{Decoherence and recoherence of a kicked beam}\label{sboscb}
Consider the case where the beam distribution is shifted by \(x_0\) uniformly
in the \(x\)-direction as an initial condition.
In practice, this is realized by giving an offset at injection or
by kicking the entire bunch with a kicker magnet located \(\pi/2\)
upstream in betatron phase.
\beq
\Psi(\bfx)=f(x-x_0,p_x)\psi(J). \label{FlatDipole}
\eeq
For \(\xi=\chi=0\), it is evident that the center of mass of the bunch
undergoes betatron oscillation with tune \(\nu_x\).
The bunch oscillation then contains only the \(l=0\) mode.
For a finite chromaticity \(\chi\ne 0\), the eigenmodes with eigenfrequencies
\(\nu_x+l\nu_z\) are \(e^{il\phi+i\chi z/\sigma_z}\),
so a uniformly shifted distribution is represented
by a superposition of multiple \(l\) modes:
\beq
x_0=\sum_l x_l(J) e^{il\phi}e^{i\chi z/\sigma_z}.\label{lximode}
\eeq
\(x_l(J)\) is obtained by applying \(\int d\phi\, e^{-il\phi-i\chi z/\sigma_z}\)
to both sides:
\beqa
x_l(J)&=&\frac{x_0}{2\pi}\int_0^{2\pi} d\phi\, e^{-il\phi}e^{-i\chi z(J,\phi)/\sigma_z}\nonumber \\
&=& x_0 i^{-l} J_l\left(\chi r\right),\label{xl0}
\eeqa
where \(r=\sqrt{2J/\varepsilon_z}\) can be viewed as a normalized radial coordinate
in the longitudinal phase space.
\(J_l(x)\) is the \(l\)-th order Bessel function.
The Bessel-function identity
\(e^{-ix\cos\phi}=\sum_{l=-\infty}^{\infty}i^{-l}J_l(x)e^{il\phi}\)
ensures that summing over all \(l\) yields the uniform shift \(x_0\).
In practice, \(l\) is truncated at some value \(l_{\max}\).
Let us examine how many \(l\) modes are needed to represent a uniform shift.
Substituting Eq.~(\ref{xl0}) into Eq.~(\ref{lximode}) and truncating at \(\pm l_{\max}\):
\beq
v(J,\phi)=x_0\sum_{l=-l_{\max}}^{l_{\max}} i^{-l}
 J_l\left(\chi r\right)
e^{il\phi+i\chi r\cos\phi}.
\eeq

\begin{figure}[tbp]
  \centering
  \includegraphics[width=80mm]{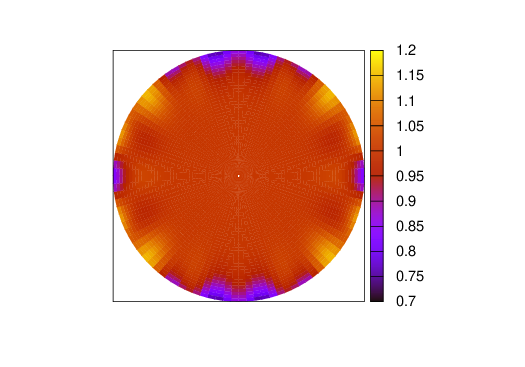}
  \caption{Reconstruction of intra-bunch phase amplitude for a uniform initial amplitude 
  with $l_{max}=8$, $\chi=8$.}
  \label{x0chi8}
\end{figure}

For zero chromaticity (\(\chi=0\)), the \(l=0\) mode alone can represent the uniform shift,
as mentioned above.
No dipole moment is associated with the oscillation of the \(l\ne 0\) modes.
For \(\chi\ne 0\), various \(l\) modes mix, because they possess a dipole mode.
Figure~\ref{x0chi8} shows the distribution for \(r<1\) in the case \(l_{\max}=8\), \(\chi=8\).
It is distributed in \(0.73<x(J,\phi)<1.15\), roughly reproducing a uniform distribution.
For \(\chi=6\), \(0.96<x(J,\phi)<1.04\), and for \(\chi=10\), \(0.17<x(J,\phi)<1.48\),
so \(l_{\max}\ge\chi\) is desirable.
This makes sense when we consider that \(\chi\) is the wavenumber in the \(z\) direction
and \(l\) is the wavenumber in the azimuthal direction.
Corresponding to the shift in \(x\), the imaginary part of \(v\), \(p_x(J,\phi)\), arises.
%

\subsubsection{Time evolution}
Let us consider the time evolution of this uniformly shifted bunch.
Each \(l\) mode moves with its own frequency (\(\nu_x+l\nu_z\)):
\beq
v(J,\phi;t)=2\pi x_0 \sum_l i^{-l} J_l\left(\chi r\right)
e^{il\phi+i\chi z/\sigma_z} e^{-2\pi i(\nu_x+l\nu_z) t}.\label{x_t}
\eeq
Here, time \(t=s/C\) is measured in turns.
At \(t=0\), \(v\) is real and only has an \(x\) component, but an imaginary part \(p_x\)
appears over time.

\begin{figure}[tbp]
  \centering
 \includegraphics[width=60mm]{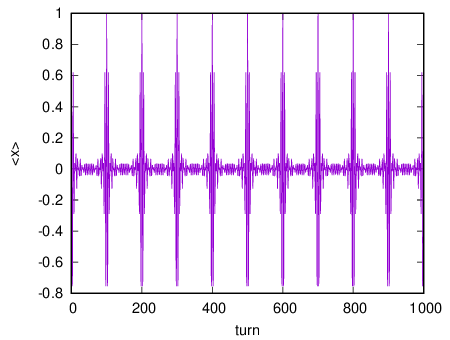} 
  \caption{ Oscillation of the bunch centroid $\langle x \rangle$ with initial amplitude 1, 
  without space charge forces.}
  \label{xchi6t1000}
\end{figure}

Figure~\ref{xchi6t1000} shows the center of mass \(\langle x \rangle\) for 1000 turns.
Even if the dipole amplitude disappears momentarily, it revives every 100 turns
(\(1/\nu_z\)).
Multiplying by the phase variation due to the tune in Eq.~(\ref{vaver}):
\beq
\langle x \rangle (t)
=x_0\sum_l \int_0^{\infty}J_l(\chi r)^2\psi(r)r\,dr\,
e^{-2\pi i (\nu_x+l\nu_z)t}.
\eeq
If the longitudinal distribution is Gaussian, \(\psi(r)=e^{-r^2/2}/(2\pi)\)
(i.e., \(\psi(J)=e^{-J/\varepsilon_z}/(2\pi\varepsilon_z)\), the integral can be performed:
\beq
\langle x \rangle (t)=e^{-\chi^2}\mathrm{Re}\left[e^{-2\pi i \nu_x t}\sum_l I_l(\chi^2)e^{-2\pi i l \nu_z t}
\right],
\eeq
where \(I_l(x)\) is the modified Bessel function.
The periodic behavior with period \(1/\nu_z\) can be inferred because the sum becomes
large when \(t\) is an integer multiple of \(1/\nu_z\).

\subsection{Frequency response to forced oscillation: Beam Transfer Function}\label{NoSCBTF}
We set up a device that generates an electromagnetic field at high frequency
at a specific location (\(s=0\)) in the accelerator, and that gives a transverse
momentum kick to the beam at a given frequency once per revolution:
\beq
F_{x,ext}(s,\omega)=A \sum_{n} e^{-i\omega (s-z)/(c\beta)}\delta(s-nC).
\eeq
A change of momentum \(dp_x=F(0,\omega)\,ds=A e^{i\omega z/(c\beta)}\) occurs at \(s=0\).
The corresponding Hamiltonian is \(H\,ds=-A e^{i\omega z/(c\beta)}x\).
The distribution function \(f\) changes as follows:
\beq
e^{:H:ds}f(x,p_x)=f(x,p_x-A e^{i\omega z/(c\beta)}).
\eeq
The average momentum \(p_x(z,\delta)\) changes by:
\beqa
p_x(z,\delta;ds)&=&\int dx\,dp_x\, p_x f(x,p_x-A e^{i\omega z/(c\beta)}) \nonumber \\
&=&p_x(z,\delta)+A e^{i\omega z/(c\beta)}.
\eeqa
To find the \(l\) component of \(\Delta p_x=Ae^{i\omega z/(c\beta)}\),
we multiply both sides by \(e^{-i\chi z/\sigma_z}e^{-il\phi}\) and integrate over \(\phi\):
\beqa
\Delta p_l(J,ds)&=& A \int \frac{d\phi}{2\pi}\, e^{-i\chi z/\sigma_z}e^{-il\phi +i\omega z/(c\beta)}\nonumber \\
&=&Ai^{l}J_l\left((\kappa-\chi)r\right).
\eeqa
Here, the input frequency is normalized by \(\kappa=\omega\sigma_z/(c\beta)\).
Since \(H\) includes \(z\), it also changes the longitudinal distribution \(\psi\),
but this is ignored since it is small.
Considering the repetition over \(s\), the change in \(v_l\) is described by
a differential equation:
\beqa
\lefteqn{\frac{dv_l}{ds}+\frac{2\pi}{C} i(\nu_x+l\nu_z)v_l} \label{deqvl}\\ &&=
 Ai^{l}J_l\left((\kappa-\chi)r\right) e^{-2\pi i n\nu}\delta(s-nC),
\nonumber
\eeqa
where \(\nu=\omega/\omega_0=\omega C/(2\pi c\beta)\) is the tune of the input frequency.
The solution of Eq.~(\ref{deqvl}) is obtained by Fourier transformation,
applying \(\int ds\, e^{2\pi i \nu' s/C}\) to both sides:
\beqa
\lefteqn{-\frac{2\pi i}{C}(\nu'-\nu_x-l\nu_z)\int ds\, e^{2\pi i \nu' s/C}v_l(J;s)}\nonumber \\
&&=Ai^l \int ds\, J_l((\kappa-\chi)r)
e^{2\pi i (\nu's/C-n\nu)}\delta(s-nC)\nonumber \\
&&=Ai^l J_l((\kappa-\chi)r)\sum_p \delta(\nu-\nu'-p).\label{ftvl}
\eeqa
Here, the identity
\beq
\sum_{n=-\infty}^{\infty} e^{2\pi in\nu}=
\sum_{p=-\infty}^{\infty}\delta (\nu-p)\nonumber
\eeq
was used.
By the inverse transform, applying \(\int d\nu'\, e^{-2\pi i \nu' s/C}\),
the amplitude \(v_l(J)\) of the \(l\) mode is obtained:
\beq
v_l(J;t)=\frac{Ai^{l+1}}{2\pi} \sum_p \frac{J_l((\kappa-\chi)r)}{\nu'-\nu'_x-l\nu_z}
e^{-2\pi i \nu t}.
\label{iftvl}
\eeq
where $\nu'_x$ is the fractional part of the tune $\nu_x$.
In the sum over integer \(p\), \(\nu'=\nu-p\) is recast by adding
the integer part of \(\nu_x\): \(p+\mathrm{int}(\nu_x)\rightarrow p\) is used as the new \(p\).
In this way, when the input frequency \(\nu\) is close to
\(\nu'_x+l\nu_z\) plus an integer \(p\), i.e., \(\nu'\sim \nu'_x+l\nu_z\),
the denominator in Eq.~(\ref{iftvl}) becomes close to zero, forming a peak.
It is common practice to perform tune measurements by applying a forced
excitation at a frequency that differs from the beam betatron frequency
by an integer multiple of the revolution frequency.

The condition for the response from modes excited by a sweep of the input frequency
(\(\nu\)) to form a peak is that the denominator of Eq.~(\ref{iftvl}) becomes zero:
\(\nu=\nu_{l,p}\equiv p+\nu'_x+l\nu_z\).
Cases with \(\nu<0\) are also considered.
The tune (frequency) at which the response peaks is \(\nu=\nu_{l,p}\),
corresponding to \(\kappa_{l,p}=\omega_{l,p} \sigma_z/(c\beta)=2\pi\nu_{l,p} \sigma_z/C\).

By calculating \(\rho_x(z)\) and \(\langle x \rangle\) for \(p\) and \(l\),
the magnitude of the response of each mode to the input frequency can be evaluated.
Let us focus on the numerator amplitude in Eq.~(\ref{iftvl}), i.e., neglect
\(A\), \(i^{l+1}\), etc., and obtain the relative amplitude for \(l,p\):
\beqa
|\langle v_{l,p} \rangle| &\sim& \int rdr\frac{d\phi}{2\pi} J_l((\kappa_{l,p}-\chi)r)
e^{il\phi+i\chi z/\sigma_z}\psi(r)\nonumber \\
&\sim& \int rdr J_l((\kappa_{l,p}-\chi)r)J_l(\chi r)
\psi(r).\label{vpl}
\eeqa

\begin{figure}[tbp]
  \centering
\includegraphics[width=60mm]{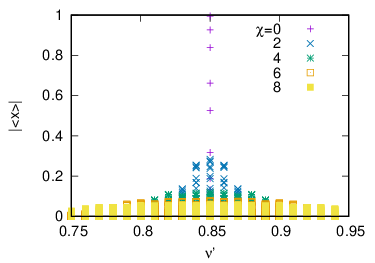} 
  \caption{Relation between output tune and amplitude for each chromaticity
  (without space-charge forces), where \(\nu'_x=0.85\) and \(\nu_z=0.01\).}
  \label{foscx0}
\end{figure}

In the case of \(\chi=0\), the amplitudes for modes other than \(l=0\) are zero
because of \(J_{l\ne0}(0)=0\).
Figure~\ref{foscx0} shows the amplitude \(|\langle v_{l,p} \rangle|\) as a function
of the fractional tune \(\nu'\) of the input frequency for each chromaticity.
The amplitude is plotted at the fractional input frequency \(\nu'=\nu'_x\pm l\nu_z\),
and those for different \(p\) are overlaid.
A symmetric amplitude distribution appears around the tune \(\nu_x=0.85\).
At \(\chi=0\), only the betatron mode \(\nu'=0.85=\nu'_x\) responds strongly
in several $p$ near $p\sim 0$.
At \(\chi>0\), the maximum is at \(\nu'=0.85\), and the response becomes weaker as
\(\chi\) increases.
The relation between tune and amplitude for \(\chi<0\) (\(\xi>0\)) is similar
to that for \(\chi>0\) (\(\xi<0\)). There are slight differences in how
the fractional part of the tune is handled.

The dipole moment as a function of \(z\) is also a measurable quantity,
\beq
\rho_{x,l}(z)\sim\int d\delta\, J_l((\kappa_{l,p}-\chi)r)\psi(J)e^{il\phi+i\chi z/\sigma_z}.
\eeq
We are interested in the Fourier component of the dipole moment at the input
frequency \(\omega=\omega_{l,p}\),
\beqa
\rho_{x}(\omega_{l,p})=\int dz\, \rho_{x,l}(z) e^{-i\omega_{l,p} z/(c\beta)}\nonumber \\
 \sim\int r\,dr\, J_l((\kappa_{l,p}-\chi)r)^2\psi(r).\label{rhoxpl}
\eeqa

\begin{figure}[tbp]
  \centering
  \includegraphics[width=60mm]{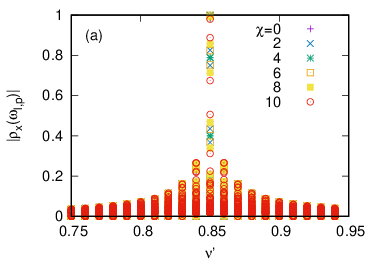} 
  \includegraphics[width=60mm]{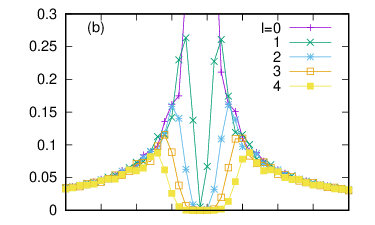}\vspace{-3mm} 
   \includegraphics[width=60mm]{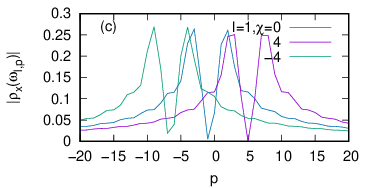} 
\caption{Beam response \(\rho_x(\omega_{l,p})\) for forced oscillation at the frequency \(\omega_{l,p}\)
  (without space-charge forces).
  (a) the response as a function of the fractional input tune \(\nu'\) for various \(\chi\),
  with different \(p\) overlaid,
  (b) the response as a function of \(p\) for each \(l\) at \(\xi=0\), and
  (c) the response for different \(\chi\) in the \(l=1\) mode.}
  \label{foscx26}
\end{figure}

Figure~\ref{foscx26} shows the Fourier component \(\rho_{x}(\omega_{l,p})\) at the input frequency.
\(\rho_{x}(\omega_{l,p})\) is plotted at the fractional input frequency
\(\nu=\nu_x\pm l\nu_z\), and those for different \(p\) are overlaid in plot (a).
For \(\chi=0\), \(\rho_{x}(\omega_{0,p})\) with \(l=0\) has a peak value \(\approx 1\) at
\(\omega_{0,-1}=(1-\nu'_x)\omega_0\). It has a decent value at $p\le 4$.
\(\rho_{x}(\omega_{l,p})\) for \(l\ge 0\) is shown in plot (b).
In Eq.~(\ref{rhoxpl}), the integral can be evaluated using the integrand
at \(r=1\): \(\rho_{x}(\omega_{l,p})\sim J_l(\kappa_{l,p}-\chi)^2\).
This behavior is consistent with plot (b).
Note that \(\Delta\kappa/\Delta p=2\pi\sigma_z/C=0.689\).

Eq.~(\ref{rhoxpl}) shows that the peak structure seen in plot (b) appears symmetrically
for $p$ determined by \(\kappa_{0,p}=\chi\).
The peak structure in plot (b) shifts by
\(\Delta p= C/(2\pi\sigma_z)\Delta\chi=1.45\Delta \chi\) as \(\chi\) is changed.
Plot (c) shows the shift for \(\chi=-4,0,4\), which gives
\(\Delta p=\pm 5\sim 6\) for \(\Delta\chi=\pm 4\).
These behaviors are reasonable, since the frequency matches the intra-bunch
variation of chromaticity \(e^{i\chi z/\sigma_z}\) when \(\kappa=\chi\).


\section{Oscillation of Beams Subject to Space-Charge Forces}\label{scosc}
\subsection{Formalism to treat space-charge force}\label{scosca}
Space-charge forces are forces acting between non-relativistically moving charged particles.
Consider how space-charge forces affect oscillations within a bunch.
Considering small oscillations of the bunch, a particle at \(x(z,\delta)\)
in phase space, or its distribution, experiences a space-charge force corresponding
to the deviation from the center $\bar{x}(z)=\int x(z,\delta)\psi(J)d\delta/\rho(z)$ integrated over 
$\delta$ at $z$:
\beq
\Delta p_x(z,\delta)=k_{sc}\beta_x\rho(z)[x(z,\delta)- \bar{x}(z)]\Delta s, \label{dpsc}
\eeq
where
\beq
k_{sc}(s)=\frac{4N r_p }{\beta^2\gamma^3}\frac{1}{R_x(R_x+R_y)}.
\label{LinSP}
\eeq

Using expressions $W_D$ and $W_Q$ for the dipolar and quadrupolar wake force integrated 
over the circumference,
\beq
W_D(z)=k_{sc}C\delta(z)\qquad W_Q(z)=-W_D(z),
\eeq
the equation of motion for the space-charge force is expressed by
\beqa
\frac{d p_{x}(z,\delta)}{ds}&=&-\frac{\beta_x}{C}
\left[\int W_Q(z-z') \rho(z')dz' x(z,\delta)\right.   \label{dpWs} \\
&+&\left.\int W_D(z-z') x(z')\rho(z')dz'\right],\nonumber
\eeqa
Impedance, which is Fourier transform of the wake force, is expressed as
\beq
Z_D=\frac{k_{sc}C}{c\beta}\equiv Z_{sc}\qquad Z_Q=-Z_{sc}.\label{Impsc}
\eeq
The impedance is inherently a function of frequency $\omega$, 
but since the wake force is a delta function, it is a constant.

Here, we transform Eq.~(\ref{dpsc}) into a form that converts a function of phase-space
variables \((z',\delta')\) into a function of \((z,\delta)\):
\beqa
\frac{d p_x(z,\delta)}{ds}&=&-\frac{\beta_x}{C}\left[x(z,\delta)
\int dz'd\delta' W_Q(z-z')\psi(z',\delta')\right. \nonumber\\
&+&\left.
\int dz'd\delta'W_D(z-z')\psi(z',\delta')x(z',\delta')\right], \label{dpscphs}
\eeqa
As in Eq. (\ref{chrmlmode}), expand $x,p_x$ into azimuthal $l$ modes.
The equation of motion for each $l$ mode is obtained by multiplying by 
$e^{-il\phi-i\chi z/\sigma_z}$ and integrating. 
\beqa
\lefteqn{\frac{dp_l(J)}{ds}=-\frac{\beta_x}{C}\int \frac{d\phi}{2\pi}} \\
&& \times \sum_{l'}\left[x_{l'}(J)e^{-i(l-l')\phi}\int dJ'd\phi' W_Q(z-z')\psi(J')\right.\nonumber\\
&& + \left.\int dJ'd\phi'W_D(z-z')\psi(J')x_{l'}(J')e^{-i(l\phi-l'\phi'+\xi (z-z')/\sigma_z)}\right], \nonumber
\eeqa
The $x(z',\delta')$ on the right-hand side is also expanded with 
$e^{il'\phi'+i\chi z'/\sigma_z}$:
Using $\delta(z)=\int d\omega/(2\pi c\beta) e^{i\omega z/(c\beta)}$ and 
the Bessel function formula:
\beqa
\lefteqn{\frac{dp_l(J)}{ds}=\frac{\beta_x Z_{sc}}{C}\sum_{l'}i^{l-l'}\left[ x_{l'}(J)
\int d\omega J_{l-l'}(\kappa r)D(\kappa)\right.}\nonumber\\
&& - \left. \int dJ'\psi(J')x_{l'}(J')\int d\omega 
J_l(\kappa r)J_{l'}(\kappa r')\right].\label{dplds} 
\eeqa

Here, variables are normalized as $\kappa=\omega \sigma_z/(c\beta)$,
$r=\sqrt{2J/\varepsilon_z}$.
$D$ in the first line of Eq.~(\ref{dplds}) is defined as:
\beq
D(\kappa)=\int^{\infty}_0 r'dr' J_0(\kappa r')\psi(r').
\eeq
The integrations have been carried out for several longitudinal distributions,
and the results for the Gaussian and water-bag distributions are summarized in
Ref.~\cite{cbisc}.
\begin{table}
  \begin{tabular}{|c|c|c|c|}
  \hline
    Type & $\psi(\hat{J})$ & $\rho(z)$ & $D(\kappa)$ \\
\hline
G & $\displaystyle \frac{1}{2\pi}e^{-\hat{J}}$ & $\displaystyle \frac{1}{\sqrt{2\pi}\sigma_z} e^{-z^2/2\sigma_z^2}$ & $\displaystyle \frac{1}{2\pi}e^{-\kappa^2/2}$\\
W & $\displaystyle \frac{1}{2\pi}h(1-\hat{J})$ & $\displaystyle \frac{1}{\pi\sigma_z^2}\sqrt{2\sigma_z^2-z^2}$ & $\displaystyle \frac{1}{\pi}\frac{J_1(\sqrt{2}\kappa)}{\sqrt{2}\kappa}$\\
\hline
\end{tabular}
  \caption{$D(\kappa)$ for several longitudinal distributions. 
  $h(x)$ is the step function. 
  The beam size is defined by $\sigma_z=\sqrt{\varepsilon_z\beta_z}$, 
  and particles are distributed over $0\le\hat{J} \le 1$ and $|z|\le \sqrt{2}\sigma_z$ 
  in the water bag distribution, where $\hat{J}=J/\varepsilon_z=r^2/2$}. \label{D}
\end{table}

In the second line of Eq.~(\ref{dplds}), \(\kappa-\chi\) enters the integrand,
but by shifting the integration over \(\omega\), it can be written in a form independent
of \(\chi\). This is possible because the impedance of the space-charge force is constant.

Here, as in Eq.~(\ref{deqvl}), we express it as an equation of motion for \(v_l\):
\beqa
\lefteqn{\frac{dv_l(J)}{ds}+\frac{2\pi}{C} i(\nu_x+l\nu_z)v_l(J)} \nonumber \\ &&=
\frac{Z_{sc}\beta_x}{2C}\sum_{l'}i^{l-l'+1}\left[ v_{l'}(J)
\int d\omega J_{l-l'}(\kappa r)D(\kappa)\right.\nonumber\\
&& - \left. \int dJ'\psi(J')v_{l'}(J')\int d\omega 
J_l(\kappa r)J_{l'}(\kappa r')\right],\label{deqvlsp}
\eeqa
Because the change due to the space-charge force is equally distributed between \(x_l\)
and \(p_l\), a factor of \(1/2\) appears.
Setting \(v_l(J)\propto e^{-2\pi i \nu s/C}\):
\beqa
\lefteqn{(\nu-\nu_x-l\nu_z) v_l(J)} \nonumber \\ &&=-
\frac{ Z_{sc}\beta_x}{4\pi}\sum_{l'}i^{l-l'}\left[ v_{l'}(J)
\int d\omega J_{l-l'}(\kappa r)D(\kappa)\right.\nonumber\\
&& - \left. \int d\omega \int dJ'\psi(J')v_{l'}(J')
J_l(\kappa r)J_{l'}(\kappa r')\right],\label{eigentunesp}
\eeqa
By solving this equation as an eigenvalue problem, the eigen-oscillation modes and
their eigenfrequencies can be found.

Eq.~(\ref{eigentunesp}) is reduced to an eigenvalue problem
by discretizing in \(J\).
Actually, we discretize with \(r=\sqrt{2J/\varepsilon_z}\)
 (\(r_k=k\Delta_r\)) instead of \(J\).
The orthogonal functions \(h_k(r)\) are expressed as
\beq
h_k(r)=1/\Delta_r \quad\mbox{for}\quad |r-r_k|\le\Delta_r/2
\eeq
and \(h_k(r)=0\) elsewhere.
The normalization and orthogonality relations are:
\beq
\int dr\, h_k(r)=1,\qquad
\int_0^{\infty} dr\, h_i(r) h_k(r)=\frac{\delta_{ik}}{\Delta_r}. \label{orthh}
\eeq
This method was first used by K.~Oide for the analysis of longitudinal instabilities
\cite{oideWz} and extended to the transverse beam-beam instabilities \cite{bbmc}.

\begin{figure}[tbp]
  \centering
  \includegraphics[width=60mm]{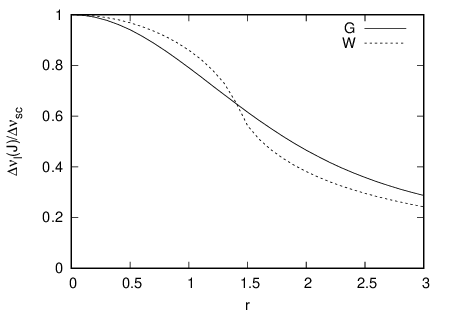} 
  \caption{Plot of \(\Delta\nu_l(r)/\Delta\nu_{sc}\) for the Gaussian and water-bag distributions.}
  \label{dnuJ}
\end{figure}

Before solving the eigenvalue problem, we try to calculate the tune shift as the
lowest-order perturbation, taking the diagonal elements of Eq.~(\ref{eigentunesp}).
As explained in detail later, taking the discretization as the basis of the
eigenvalue problem, the first term of the diagonal element on the right side of
Eq.~(\ref{eigentunesp}) is larger than the second term.
The \(J\)-dependent tune shift is:
\beq
\Delta \nu_{x,l}(r)=-\frac{k_{sc}\beta_x C}{4\pi\sigma_z}
\int^{\infty}_{-\infty} d\kappa\, J_0(\kappa r)D(\kappa)\label{TuneShiftJ}
\eeq
The tune shift, which is independent of \(l\), is given for the Gaussian distribution as
\beq
\Delta\nu_l(r)=\Delta\nu_{sc} e^{-r^2/4}
I_{0}\left(\frac{r^2}{4}\right), \quad \Delta\nu_{sc}=
-\frac{\beta_x}{4\pi}\frac{k_{sc}C}{\sqrt{2\pi}\sigma_z}.\label{TSG}
\eeq
That for the water-bag distribution is
\beqa
\Delta\nu_l(r)=\Delta\nu_{sc} \int^\infty_0 J_0(\kappa r)
J_1(\sqrt{2}\kappa)\frac{d\kappa}{\kappa},\label{TSW}\\
\Delta\nu_{sc}=-\frac{\beta_x}{4\pi}\frac{\sqrt{2}k_{sc}C}{\pi\sigma_z}.\nonumber
\eeqa
The coefficient \(\Delta\nu_{sc}\) is the so-called space-charge tune shift at the peak
line density \(\rho(0)\).
Figure~\ref{dnuJ} shows the \(r\)-dependent term, \(\Delta\nu_l(r)/\Delta\nu_{sc}\),
given in Eqs.~(\ref{TSG}) and (\ref{TSW}).

Next, let us move on to solving the eigenvalue problem.
Expand \(v_x(J,\phi)\) in terms of \(h_k(r)\) and represent the coupling coefficient
as \(v_{lk}\):
\beq
v_l(J)=\sum_{k} v_{lk}h_k(r) \label{radial} 
\eeq
Applying \(\Delta_r \int dr\, h_k(r)\) to Eq.~(\ref{eigentunesp}) yields:
\beqa
\lefteqn{(\nu-\nu_x-l\nu_z) v_{lk} }\nonumber \\ &&=
-\frac{ Z_{sc}\beta_x}{4\pi}\sum_{l'}i^{l-l'}\left[ v_{l'k}
\int d\omega\, J_{l-l'}(\kappa r_k)D(\kappa)\right.\nonumber\\
&& - \left. \sum_{k'} r_{k'}\psi(r_{k'})\Delta_r v_{l'k'}
\int d\omega\,
J_l(\kappa r_k)J_{l'}(\kappa r_{k'})\right],\label{eigeneqlk}
\eeqa
With the change of variables from \(J\) to \(r\), the distribution function will
henceforth be written as \(\psi(r)\), with the two related by \(\psi(J)dJ=\psi(r)r\,dr\).

The eigenvalue equation to be solved is:
\beq
\left(\nu_x+l\nu_z+M\right) \bfv =\nu \bfv. \label{TuneShift}
\eeq
The first two terms on the left side are the diagonal matrix
\((\nu_x+l\nu_z)\delta_{ll'}\delta_{kk'}\),
and the third term, the matrix \(M\), is expressed as:
\beqa
\lefteqn{M_{lkl'k'}=-
\frac{Z_{sc}\beta_x}{4\pi}i^{l-l'}
\left[\delta_{kk'}
\int d\omega\, J_{l-l'}(\kappa r_k)D(\kappa)\right.}\nonumber\\
&& -  \left. r_{k'}\psi(r_{k'})\Delta_r \int d\omega\,
J_l(\kappa r_k)J_{l'}(\kappa r_{k'})\right], \label{Mlk}
\eeqa
Equations~(\ref{TuneShift}) and (\ref{Mlk}) constitute a complex square matrix of
dimension \((2l_{\max}+1)n_J\).
Here, we introduce a cutoff for the frequency in the integral. 
Since $r$ was discretized with $\Delta r$, we used $\kappa_c=1/\Delta r$, 
i.e. $\omega_c=\sqrt{2}c\beta/(n_J\sigma_z)$, 
as the cutoff frequency and set it ($\pm\omega_c$) as the upper and lower limits of the $\omega$ integral.

Since this matrix is not Hermitian, the eigenvectors satisfy the following orthogonality
relation, represented by left and right eigenvectors:
\beq
\sum_{lk}v_{lk}^{(m,L)*}v_{lk}^{(m',R)}=\delta_{mm'}.\label{orthm}
\eeq
The eigen-oscillation states are represented by a linear combination of
the basis vectors consisting of the radial function \(h_k(r)\) and
the azimuthal function \(e^{il\phi+i\chi z/\sigma_z}\):
\beqa
v(J,\phi)_m&=&\sum_{lk} v_{lk}^{(m,R)} h_k(r) e^{il\phi+i\chi z/\sigma_z}\label{vm}\\
v(J,\phi)_m^{\dag}&=&\sum_{lk} v_{lk}^{(m,L)*} h_k(r) e^{-il\phi-i\chi z/\sigma_z}.
\eeqa
The state corresponding to the conjugate is defined as \(v(J,\phi)_m^{\dag}\).
The orthogonality relation for the eigen-oscillation states is:
\beq
\frac{1}{2\pi}\int dr\,d\phi\, v(J,\phi)_{m'}^{\dag}v(J,\phi)_m=\frac{\delta_{mm'}}{\Delta_r}.
\eeq
Here, orthogonality in \(l\) and the orthogonality relations for \(k,m\),
Eqs.~(\ref{orthh}) and (\ref{orthm}), were used.
The time evolution of the oscillation state involves a phase rotation by
the eigenfrequency (tune) at each turn:
\beq
v(J,\phi;t)_{m}=e^{-2\pi i\nu_m t} v(J,\phi;0)_{m}.
\eeq
Any oscillation state can be represented by a linear combination of eigenvectors:
\beq
v(J,\phi)=\sum_m a_m v(J,\phi)_m
\eeq
Given an oscillation state \(v(J,\phi)\), the coefficients of the linear combination
\(a_m\) can be determined by applying \((2\pi)^{-1}\int dr\,d\phi\, v(J,\phi)_m^{\dag}\)
to both sides:
\beqa
\lefteqn{a_m=\frac{\Delta_r}{2\pi} \int dr\, d\phi\, v(J,\phi)_m^{\dag}v(J,\phi)}\nonumber\\
&&=\frac{\Delta_r}{2\pi} \int d\phi \sum_{lk} v_{lk}^{(m,L)*}
e^{-il\phi-i\chi z_k/\sigma_z}v(J_k,\phi).\label{am}
\eeqa


\subsection{Beam oscillation mode given by solving the eigenvalue problem}\label{scoscb}
The eigenvalue problem was solved using the parameters of the CSNS,
where the longitudinal phase-space distribution was Type W in Table~\ref{D}.
This is a uniform-density distribution in phase space, a so-called water-bag distribution.
Using a vector of length \(40\times 41=1640\), with \(n_J=40\), \(l_{\max}=20\),
we solve the eigenvalue problem for a \(1640\times 1640\) matrix for each bunch intensity.

\begin{figure}[tbp]
\includegraphics[width=60mm]{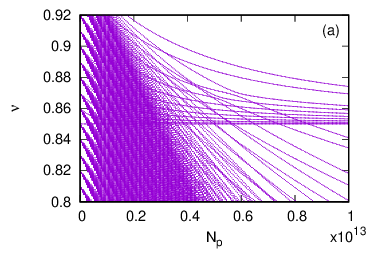} 
\includegraphics[width=60mm]{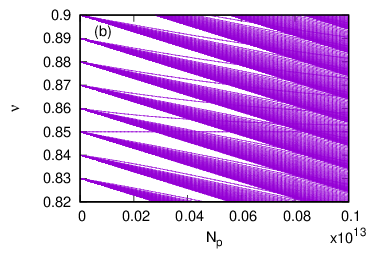} 
\caption{Tunes of oscillations for each bunch intensity, obtained by solving
  the eigenvalue problem, where \(l_{\max}=10\) and \(n_J=40\).
  The horizontal scale of the bunch population
  is \(N_p=0\)--\(1\times 10^{13}\) in (a) and is enlarged to \(0\)--\(1\times 10^{12}\) in (b).}
  \label{scmode}
\end{figure}

Figure~\ref{scmode} plots the eigen-tunes as a function of the bunch population, where
the number of particles in the bunch was varied up to \(N_p=1\times 10^{13}\) in 100 steps,
and the eigenvalue problem was solved at each step.
Without space-charge forces, the eigen-tunes were \(\nu_x+l\nu_z\) and were degenerate
for the radial modes \(v_l(J)\),
but the space-charge force lifts this degeneracy, generating a tune shift for each radial mode.
Particles with smaller synchrotron amplitudes experience a larger negative tune shift
due to stronger space-charge forces, and vice versa.
Many of the lines shifting negatively with increasing \(N_p\) in the figure are due to
the space-charge tune shift.
The magnitude of the shift corresponds to the synchrotron amplitude (\(r\) or \(J\)).

In Fig.~\ref{scmode}, for \(l\ge 0\), there are modes clearly isolated from other modes.
For better readability of the figure, $l_{max}$ is set to 10.
This isolated mode is easily understood. When the chromaticity is 0,
a distribution where all particles in the bunch are uniformly shifted undergoes
betatron oscillations without being affected by space-charge forces.
For \(l<0\), it is not clear whether there are isolated modes.

\begin{figure}[tbp]
  \centering
   \hspace*{-1.cm}
   \includegraphics[width=40mm]{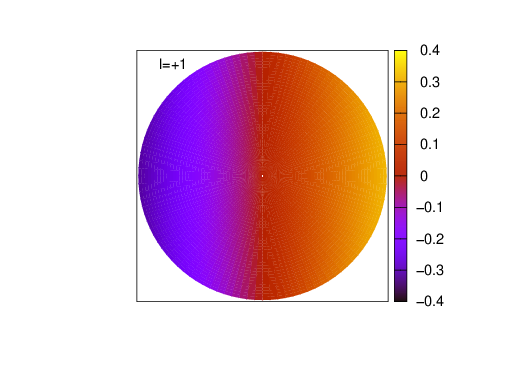}    \hspace*{-1.5cm} 
  \includegraphics[width=40mm]{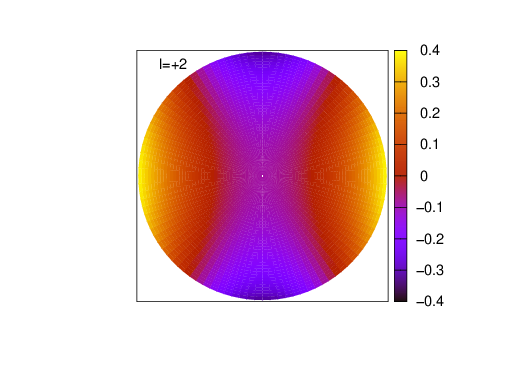}    \hspace*{-1.5cm}
  \includegraphics[width=40mm]{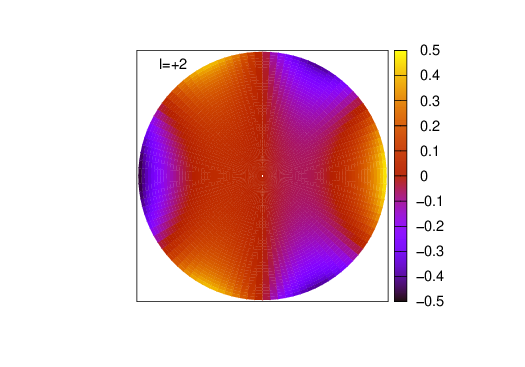}    \hspace*{-2.cm}
\includegraphics[width=40mm]{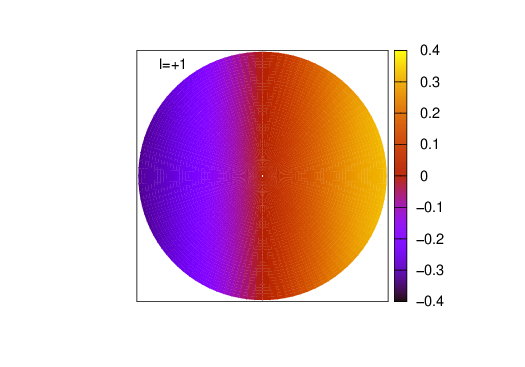}  \hspace*{-1.5cm}
  \includegraphics[width=40mm]{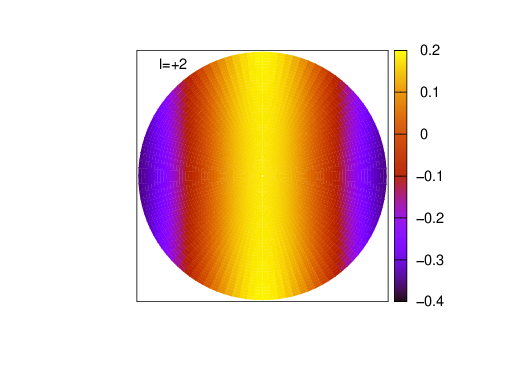}   \hspace*{-1.5cm}
  \includegraphics[width=40mm]{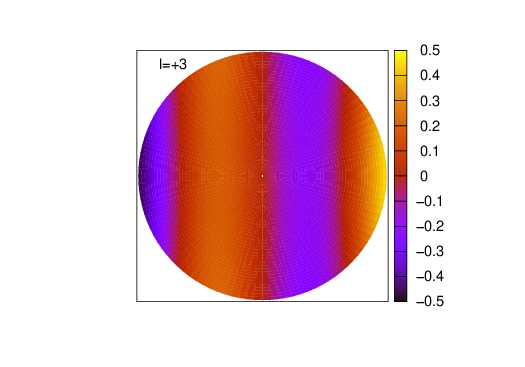}   \hspace*{-1.5cm}
   \caption{Eigen-oscillation patterns of the separated modes obtained from Eq. (\ref{vm}) 
  using the eigenvectors.
  Horizontal and vertical axes are $z$ and $\delta$, respectively.
  Top row: $N_p=0.44\times 10^{12}, |\Delta\nu_{sc}/\nu_z|=1.6$, 
  Bottom row: $N_p=3.5\times 10^{12}, |\Delta\nu_{sc}/\nu_z|=12$.
  From left to right: $l=1,2,3$.}
  \label{eigenvectors}
\end{figure}

Figure~\ref{eigenvectors} shows the eigen-oscillation patterns of the separated modes
represented by Eq.~(\ref{vm}).
The top row is for \(N_p=0.44\times 10^{12}\), arranged from left to right for \(l=1,2,3\).
Only the real part is plotted, so the amplitude distribution \(x(J,\phi;0)\) is depicted.
The distributions show vertical oscillations corresponding to \(l=1,2,3\)
in the azimuthal direction.
Tracking the time evolution by multiplying by \(e^{-i\omega t}\) reveals
the oscillation pattern rotating in the \(z\)-\(\delta\) phase space due to
synchrotron oscillations.

The bottom row shows the oscillation patterns for \(N_p=3.5 \times 10^{12}\).
They are uniform in the \(\delta\) direction, and the structure is only seen in the \(z\) direction.
Space-charge forces do not act on such a distribution as seen in Eq.(\ref{dpsc}).
Individual particles undergo synchrotron oscillations.
Space-charge forces act so that synchrotron oscillations are not coherently visible
in the direction of synchrotron motion.
When tracking the time evolution, synchrotron oscillations are not seen when the amplitude
is large, but rotation becomes visible when the amplitude becomes small.
The tune approaches \(\nu_x\) but does not completely reach \(\nu_x\),
so the synchrotron oscillation component has not vanished.

\subsection{Decoherence and recoherence of a kicked beam}\label{scoscc}
Consider a distribution uniformly shifted in the \(x\) direction. The initial state is
given by \(v(J,\phi;0)=x_0\).
This initial distribution is expanded in terms of the eigenvectors:
\beq
x_0=\sum_m a_m v(J,\phi)_m.
\eeq
Substituting into Eq.~(\ref{am}), the coefficients \(a_m\) are obtained for \(x_0=1\) as:
\beq
a_m=\Delta_r\sum_{lk} v_{lk}^{(m,L)*} i^{-l} J_l\left(\chi r_k\right). \label{init_amp}
\eeq

\begin{figure}[tbp]
  \centering
    \includegraphics[width=60mm]{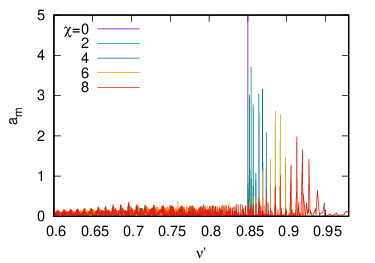} 
  \caption{Linear combination coefficients \(a_m\) for the beam initialized by a uniform dipole
  distribution, obtained from Eq.~(\ref{init_amp}).}
  \label{am_sc2}
\end{figure}

The extent to which each eigenmode is included in the uniformly shifted distribution
can be seen from \(a_m\).
Figure~\ref{am_sc2} shows the amplitudes of the eigenmodes for each chromaticity \(\chi\).
At \(\chi=0\), only the \(l=0\) mode with tune equal to \(\nu_x\) is induced.
For \(\chi>0\), as \(\chi\) increases, modes with tunes higher than \(\nu_x\) become included.
Modes with tunes lower than \(\nu_x\) are continuously distributed as seen in Fig.~\ref{scmode},
so individual modes have small components.

\begin{figure}[tbp]
  \centering
\includegraphics[width=60mm]{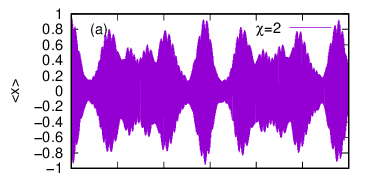}\vspace{-3mm} 
  \includegraphics[width=60mm]{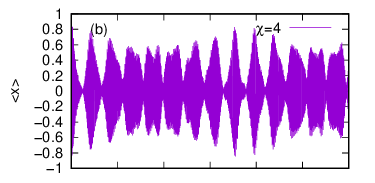}\vspace{-3mm}
  \includegraphics[width=60mm]{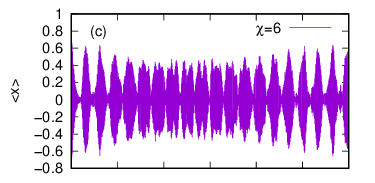}\vspace{-3mm}
  \includegraphics[width=60mm]{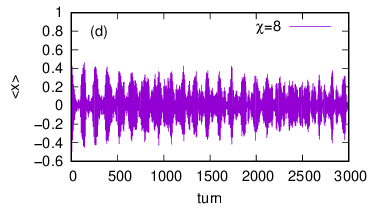}
  \caption{Centroid oscillation of the bunch for each chromaticity, with an initial amplitude
  of \(x_0=1\), obtained from Eq.~(\ref{xav_t}), where \(N_p=3.5\times 10^{12}\), 
  \(|\Delta\nu_{sc}/\nu_z|=12\).}
  \label{evtr_csns10}
\end{figure}

Each eigenmode oscillates according to its own eigen-tune \(\nu_m\):
\beq
v(J,\phi;t)=\sum_m a_m v(J,\phi)_m e^{-2\pi i\nu_m t}.\label{xav_t}
\eeq
Figure~\ref{evtr_csns10} plots the centroid oscillation of the bunch for each chromaticity.
Coherence is lost due to differences in the tunes of the included eigenmodes,
and the centroid initially approaches zero, but coherence recovers again.
The recovery period is \(1/\nu_z=100\) turns, as seen in Fig.~\ref{xchi6t1000}
without the space-charge force, but with the space-charge force, the period becomes longer than
\(1/\nu_z\).
Although the \(l=1\) mode is quite close to the tune, the modes included vary
with chromaticity as seen in Fig.~\ref{am_sc2}, so the degree of deviation of
the period from \(1/\nu_z\) depends on chromaticity.
For small chromaticity, a small number of modes \(\nu_x<\nu<\nu_x+\nu_z\) are excited, so
the periodicity is clearly longer than \(1/\nu_z\).
For high chromaticity, several to tens of modes with \(\nu>\nu_x\) are excited, so the behavior is
complex, though the periodicity is longer than \(1/\nu_z\).

\subsection{Frequency response of forced oscillation: Beam Transfer Function}\label{SCBTF}
For the case with a strong space-charge force, we find the response of each eigenmode to
a sinusoidal forced oscillation \(\Delta p_x(J,\phi)=Ae^{i\omega z/(c\beta)}\).
The momentum change at \(s=0\) is given by
\beq
\Delta v_x(J,\phi)=\sum_m \Delta a_m v(J,\phi)_m=iAe^{i\omega z/(c\beta)}.
\eeq
Multiplying by \(v(J,\phi)_m^\dag\) and integrating over \(J,\phi\),
the coefficient \(\Delta a_m\) for each eigenmode is obtained:
\beqa
\Delta a_m&=&\frac{i}{2\pi}\int dr \int d\phi\, A v(J,\phi)_m^\dag e^{i\omega z/(c\beta)} \nonumber\\
&=&\frac{iA\Delta_r}{2\pi}\int d\phi \sum_{lk}v_{lk}^{(m,L)*}e^{-il\phi-i\chi z_k/\sigma_z+i\omega z_k/(c\beta)}\nonumber\\
&=&A\Delta_r \sum_{lk}v_{lk}^{(m,L)*}i^{l+1}J_l((\kappa-\chi)r_k).
\eeqa
Considering the repetition over \(s\), we express the change in \(a_m\) by
a differential equation:
\beqa
\lefteqn{\frac{da_{m}}{ds}+\frac{2\pi i}{C} \nu_m a_m} \label{deqam}\\ &&=
A\Delta_r \sum_{lk} i^{l+1}v_{lk}^{(m,L)*} J_l\left((\kappa-\chi)r_k\right) \sum_n e^{-2\pi i n \nu}\delta(s-nC)
\nonumber
\eeqa
The solution of Eq.~(\ref{deqam}) is obtained similarly to
Eqs.~(\ref{ftvl})--(\ref{iftvl}), giving:
\beq
a_m(\nu)=-A\Delta_r \sum_p \frac{\sum_{lk}i^l v_{lk}^{(m,L)*}
J_l\left((\kappa-\chi)r_k\right)}{\nu'-\nu_m}e^{-2\pi i \nu't}.
\label{iftam}
\eeq
When the beam receives the momentum change at a certain frequency,
the strength of each induced mode is indicated by \(a_m\).
The condition for the linear combination coefficient \(a_m\) corresponding to
the modal response to form a peak is that the denominator of Eq.~(\ref{iftam}) becomes zero:
 \(\nu'-\nu_m=0\).
The condition means that excitation occurs when the oscillation mode tune \(\nu_m\)
and the forced oscillation tune \(\nu\) differ by an integer \(p\).
As is done in Sec.~\ref{NoSCBTF}, we express \(\nu_m\) using the fractional part.
Consider the case where a certain mode \(m\) is excited by a sweep of the input frequency
(\(\nu\)).
The tune (frequency) at which \(a_m(\nu)\) peaks corresponds to multiple integer values
\(p\), with frequencies \(\nu_{m,p}=\nu_m+p\) and \(\kappa_{m,p}=2\pi\nu_{m,p} \sigma_z/C\).
Denote the response at a specific \(p\) in the sum over \(p\) in Eq.~(\ref{iftam}) as
\(a_{m,p}\):
\beq
a_{m,p}\equiv \sum_{lk}i^l v_{lk}^{(m,L)*} J_l\left((\kappa_{m,p}-\chi)r_k\right).
\label{iftam2}
\eeq

\begin{figure}[tbp]
  \centering
\includegraphics[width=60mm]{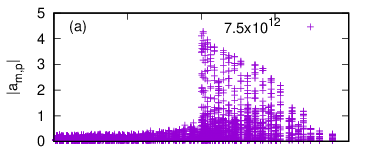}\vspace{-3mm} 
\includegraphics[width=60mm]{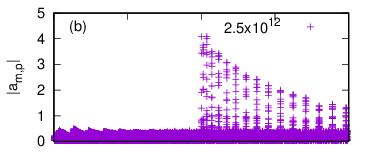}\vspace{-3mm}
\includegraphics[width=60mm]{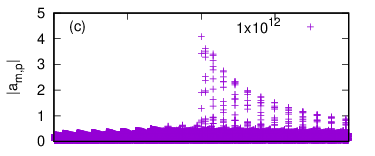}\vspace{-3mm}
\includegraphics[width=60mm]{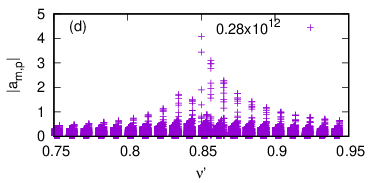}
  \caption{Amplitude \(|a_m|\) induced by forced oscillation,
  where the beam intensity is \(N_p=(7.5, 2.5, 1.0, \text{and } 0.278)\times 10^{12}\)
  (\(\Delta \nu_{sc}=0.27, 0.09, 0.036, 0.01\), and \(\nu_z=0.01\)) in plots (a)-(d).}
  \label{am_bt}
\end{figure}

Figure~\ref{am_bt} presents the amplitude \(a_{m,p}\) induced by forced oscillation for \(\chi=0\),
where the horizontal axis is the fractional part of the oscillation tune and different \(p\) are overlaid.
Points are located at \(\nu_m\) on the horizontal axis.
The eigenvectors are independent of \(\chi\) and \(p\).
As seen in Eq.~(\ref{iftam2}), \(a_m\) depends on \(\chi\) and \(p\) through the argument of
the Bessel function, \((\kappa_{m,p}-\chi)r_k\).
The figure does not change for a finite chromaticity except for a shift in \(p\),
\(\Delta p=C\chi/(2\pi\sigma_z)=\xi/\eta_p\).

The mode induced at the frequency corresponding to $\nu_{m,p}$, combined with the eigenvector, 
has the following dipole amplitude distribution:
\beq
a_{m,p} v(J,\phi)_m=a_{m,p}\sum_{lk}v_{lk}^{(m,R)}h_k(r)e^{il\phi+i\chi z/\sigma_z}.
\label{vpbtf}
\eeq
By integrating Eq. (\ref{vpbtf}) over $\delta$ or $(z,\delta)=(J,\phi)$ for 
each $p$ and $m$, measurable quantities, $\rho_x(z)$ and $\langle v\rangle$ are obtained. 
They characterize the magnitude of the response of each mode to the input frequency.
The response for $\langle v\rangle$, in which mode $m$ excited by frequency $\nu=\nu_m+p$, is:
\beqa
\lefteqn{\langle v_{m,p} \rangle = 
a_{m,p}\int rdr d\phi \psi(r) v(J,\phi)_m }\nonumber
\\ &&=2\pi a_{m,p} \sum_{lk}i^{l} v_{lk}^{(m,R)} J_l(\chi r_k)r_k\psi(r_k).\nonumber
\eeqa

\begin{figure}[tbp]
  \centering
  \includegraphics[width=60mm]{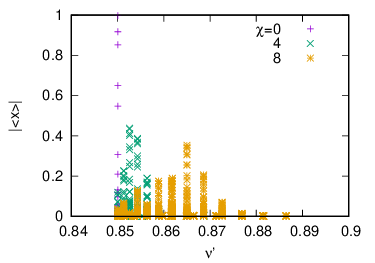} 
  \caption{Beam response for \(\langle v_{m,p} \rangle\) induced by a forced oscillation
  with frequency \(\nu_{m,p}\).}
  \label{xavnu_bt}
\end{figure}

Figure~\ref{xavnu_bt} shows the response for \(\langle v_{m,p} \rangle\) at \(N_p=7.5\times 10^{12}\).
At \(\chi=0\), the result is almost the same as without the space-charge force.
Only the betatron mode \(\nu'=0.85=\nu'_x\) responds strongly for several \(p\) near \(p=0\).
No dipole moment is associated with the oscillation of the \(l\ne 0\) modes even in the presence of
the space-charge force.
The difference is clear for \(\chi>0\) when compared to Fig.~\ref{foscx0} without the space-charge force.
A peak appears at \(l>0\) for \(\chi>0\), and the peak shifts to higher \(l\) for increasing \(\chi\).
This behavior is the same for \(\chi<0\): the peak structure depends only on \(|\chi|\).
The height of the response is higher than that without the space-charge force.
There is also no response for the \(l<0\) modes, as seen in Figs.~\ref{am_sc2} and \ref{am_bt}.

The dipole moment as a function of \(z\) is given by the integral of Eq.~(\ref{vpbtf}) over \(\delta\),
\beq
\rho_x(z)_{m,p}=a_{m,p}\int d\delta\, v(J,\phi)_m \psi(J).
\eeq
\(v(J,\phi)_m\) and its \(\delta\) integral, \(\rho_x(z)_m\), exhibit a sinusoidal structure in \(z\)
due to the chromatic modulation \(e^{i\chi z/\sigma_z}\) and the oscillation mode subject to
the space-charge force, as shown in Fig.~(\ref{eigenvectors}).
The Fourier component of the dipole moment at the input frequency \(\omega=\omega_{m,p}\) is given by
\beqa
\lefteqn{\rho_x(\omega_{m,p})=\int dz\, \rho_x(z)_{m,p} e^{-i\omega_{m,p} z/(c\beta)}}\nonumber \\
&&=a_{m,p}\int r\,dr\,d\phi \sum_{lk}v_{lk}^{(m,R)}h_k(r) \psi(J)e^{il\phi+i\chi z/\sigma_z-i\omega_{m,p} z/(c\beta)} \nonumber\\
&&= 2\pi a_{m,p}\sum_{lk} i^{-l}v_{lk}^{(m,R)}J_l\left((\kappa_{m,p}-\chi)r_k\right)r_k\psi(r_k).\label{Frhox}
\eeqa

\begin{figure}[tb]
  \centering
\includegraphics[width=60mm]{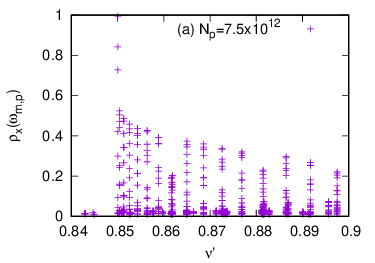} 
  \includegraphics[width=60mm]{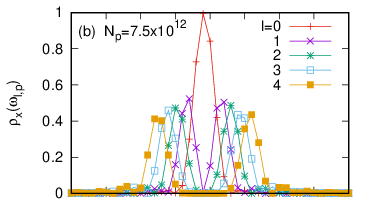}\vspace{-3mm} 
  \includegraphics[width=60mm]{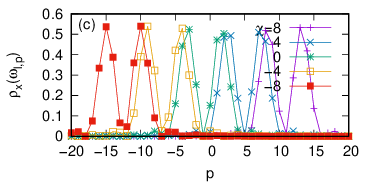} 
 \caption{Response for forced oscillation of the beam subject to the space-charge force.
  (a) \(\rho_x(\omega_{m,p})\), where those for different \(p\) are overlaid.
  (b) \(\rho_x(\omega_{m,p})\) for the isolated modes with different \(l\).
  (c) Shift of the response \(\rho_x(\omega_{m,p})\) for different chromaticities.}
  \label{rhox_bt}
\end{figure}

Figure~\ref{rhox_bt} shows the response of \(\rho_x(\omega_{m,p})\) at \(N_p=7.5\times 10^{12}\).
The corresponding figure without the space-charge force is shown in Fig.~\ref{foscx26}.
Plot (a) depicts the response at the eigen-tune \(\nu_m\), and those for different \(p\) are overlaid.
The plot is drawn for \(\chi=0\); the same plot applies for \(|\chi|\ne 0\).
The discrete mode with visible amplitude corresponds to the isolated mode seen in Fig.~\ref{scmode}.
The mode \(\nu=0.85\) is the isolated pure dipole mode. The modes \(\nu>0.85\) are lined up from
left to right as \(l=1,2,3,\ldots\). We express the isolated modes using \(l\).

\(\rho_x(\omega_{m,p})\) for the \(l\ge 0\) isolated modes is shown in plot (b), where \(\chi=0\).
As seen in Fig.~\ref{foscx26}, plot (c) shifts as \(\chi\) changes, by 
\(\Delta p=C \Delta \chi/(2\pi\sigma_z)=\Delta\xi/\eta_p\).
This is because Eq.~(\ref{rhoxpl}) contains \(J_l(\kappa_{m,p}-\chi)^2\).
Here, both Eqs.~(\ref{iftam2}) and (\ref{Frhox}) contain \(J_l(\kappa_{m,p}-\chi)\).
The eigenvectors \(v_{lk}^{(m)}\) do not depend on \(\chi\), because the space-charge impedance
in Eq.~(\ref{Impsc}) is independent of frequency.
The peak structure of \(\rho_x(\omega_{m,p})\) only shifts with \(p\), but its values do not change.
This explains why the same plot is obtained for \(|\chi|\ne 0\) in plot (a).
Plot (c) shows the shift of the response \(\rho_x(\omega_{m,p})\) for different chromaticities.

The difference from the response without the space-charge force is
(i) there is a response for only one mode for the \(l\ge 0\) modes, but no response for the \(l<0\) modes,
(ii) the frequency separation of the response between \(l\) modes is narrower than \(\nu_z\), and
(iii) the response for the \(l>0\) modes is more than twice as strong as that without the space-charge force.

This behavior is closely connected to the oscillation mode with and without the space-charge force.
(i) and (ii) are explained by the fact that an isolated mode exists for each of the \(l\ge 0\) modes,
whereas no isolated modes exist for \(l<0\).
The modes with \(l<0\) have only a continuous spectrum,
as shown in Fig.~\ref{scmode}.
(iii) is explained by the mode pattern seen in Fig.~\ref{eigenvectors}.
The distribution of the transverse amplitude \(\sum_{lk} v_{lk}^{(m)} J_l(\kappa r_k)e^{il\phi}\)
is sinusoidal in \(z\), and is independent of \(\delta\).
The Fourier integral therefore gives a larger value compared with the pure synchro-betatron mode,
\(e^{il\phi}\).

\begin{figure}[tb]
  \centering
  \includegraphics[width=60mm]{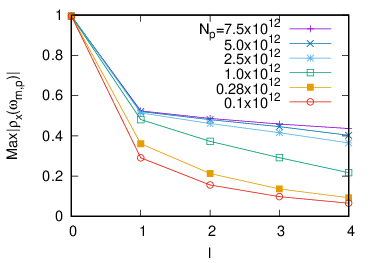} 
 \caption{Frequency response \(Max(\rho_x(\omega_{m,p}))\)
for the isolated \(l\) modes for various
bunch intensities, \(N_p=(7.5, 5.0, 2.5, 1.0, 0.28, 0.1)\times 10^{12}\),
where \(-\Delta \nu_{sc}=\nu_z=0.01\) at \(N_p=0.28\times 10^{12}\).}
  \label{rhox_summary}
\end{figure}

The continuous modes do not respond to forced oscillations: this is due to Landau damping.
Even for the \(l>0\) modes, the response is limited to the discrete modes.
Only isolated modes separated from the continuum respond to forced oscillation.
Figure~\ref{rhox_summary} summarizes the frequency response \(\rho_x(\omega_{m,p})\)
for the isolated \(l\) modes for various bunch intensities,
where \(|\Delta \nu_{sc}|=\nu_z=0.01\) at \(N_p=0.28\times 10^{12}\).
The lines with \(|\Delta\nu_{sc}/\nu_z|\ge 9\) are almost overlapping, and those
with \(|\Delta\nu_{sc}/\nu_z|\le 1\) are almost overlapping.
\(|\Delta\nu_{sc}/\nu_z|\sim 3\) shows transient behavior.
For \(|\Delta\nu_{sc}/\nu_z|\ge 9\) and \(\le 1\), the difference in the response gets bigger as \(l\) increases,
and at \(l = 3\) and 4, it differs by a factor of five.

\section{Beam Instability} \label{scinstab}
We discuss the beam instability that arises when the frequency components possessed
by the beam match those of the impedance.
Frequency shifts caused by the impedance are assumed to be small: that is, transverse mode coupling is
not considered.
The main subjects are the head-tail instability and the coupled-bunch instability.
The impedance is expressed as a resonator model with \(Q\gtrsim 10\),
and the contribution of the kick factor is small
relative to the synchrotron tune. Cases where the resonator frequency is high
with a large \(R_s\) but not a large \(R_s/Q\) are also included.
This would apply to the impedance and beam instability in high-intensity proton accelerators
such as neutron sources.
The impedance induced by the ceramic chambers in CSNS is \(Z_{\mathrm{peak}}=1~\mathrm{M}\Omega/\mathrm{m}\) at \(f_R=123\)~kHz with \(Q=40\)
~\cite{LHuang2024}.
The impedance induced by the kicker magnets in J-PARC is \(Z_{\mathrm{peak}}\sim 160~\mathrm{k}\Omega/\mathrm{m}\) at \(f_R=0.4\)--\(4\)~MHz
with \(Q=30\sim 50\)~\cite{shobuda2017}.

In conventional instability theories, the underlying oscillation modes were
those of the free synchro-betatron oscillation in Eq.~(\ref{chrmlmode}).
The synchro-betatron modes are degenerate in the radial direction,
so there is uncertainty about which radial mode to use as the base.
Generally, the lowest-order flat radial mode is used.
It has been found that the oscillation modes differ significantly
for beams with strong space-charge forces.
The eigenvalue problem including space-charge forces and the impedance was
solved in Ref.~\cite{cbisc}, where the impedance was evaluated at \(\nu_x\).
This implicitly assumed \(Q< \nu_R/\nu_z\), where \(\nu_R=f_R/f_0\) was the resonator frequency
normalized by the revolution frequency.
Here, following the conditions described earlier,
we take the underlying eigenmodes as oscillation modes that account for the space charge.
The space-charge force dominates the effects on intra-bunch oscillations,
and the instability is caused by the impedance under resonance conditions in which the oscillation mode frequency
and the impedance frequency match.

The wake force \(W_x(z)\) is now added to the equation
of motion. 
In this section, we use the wake and impedance in conventional units
$W_x[\mathrm{m}^{-2}]$ and $Z_x[\mathrm{s/m}^2]$, respectively \cite{achao}.
The equation is expressed in terms of its Fourier transform, the impedance \(Z_x(\omega)\).
\beqa
\lefteqn{\frac{dv(J,\phi;s)_m}{ds}+\frac{2\pi i \nu_m}{C} v(J,\phi;s)_m =-\frac{iN_pr_p\beta_x}{2\gamma C}
\sum_{n=-\infty}^{\infty}}\nonumber \\
&& \int W_x(z-z'-nC)v(J',\phi';s)_m e^{2\pi i n\nu_m}\psi(J')dJ'\,d\phi'\nonumber
\eeqa
Using \(v(J,\phi;s)_m=e^{-2\pi i \nu s/C}v(J,\phi;0)_m\), multiplying both sides by \(v(J,\phi)_m^{\dag}\)
and integrating over \((r,\phi)\):
\beqa
\lefteqn{(\nu-\nu_m) =-\frac{iN_pr_p\beta_x}{4\pi\gamma} \omega_0 }\label{nuinst1}\\
&& \sum_{p=-\infty}^{\infty}   Z(\omega_{mp}) g_{m,L}^*(\kappa_{mp}-\chi) g_{m,R}(\kappa_{mp}-\chi)\nonumber
\eeqa
where
\beqa
g_{m,L}^*(\kappa)&=&\sum_{lk} i^{l}v_{lk}^{(m,L)*}J_l(\kappa r_k), \nonumber\\
g_{m,R}(\kappa)&=&\sum_{lk} i^{-l}v_{lk}^{(m,R)}J_l(\kappa r_k) r_k\psi(r_k)\Delta_r.\nonumber
\eeqa
The growth rate of mode \(m\) per revolution can be determined by examining the imaginary part of
the tune multiplied by \(2\pi\):
\beqa
\lefteqn{G_r=-\frac{N_pr_p\beta_x}{2} \omega_0}\label{Gr}\\
&&\mathrm{Re}\left[\sum_{p} Z_x(\omega_{m,p})g_{m,L}^*(\kappa_{m,p}-\chi)
g_{m,R}(\kappa_{m,p}-\chi)\right]\nonumber
\eeqa
\(g_{m,L}^*(\kappa_{m,p}-\chi)g_{m,R}(\kappa_{m,p}-\chi)\) in Eq.~(\ref{Gr}) contains the same terms as
\(\rho_x(\omega_{m,p})\) expressed by Eqs.~(\ref{iftam2}) and (\ref{Frhox}).
Thus, a rough sketch of the instability can be understood from Figs.~\ref{foscx26}, \ref{rhox_bt}
and \ref{rhox_summary}.
For example, for Re\([Z(\omega)]=1~\mathrm{M}\Omega/\mathrm{m}\), the growth rate and time are \(5\times 10^{-4}\) and 4~ms,
when \(\nu_R=-\nu_x+p\). The growth rate is positive when the impedance is negative: that is, \(\omega<0\).
The condition and the result do not change with and without the space-charge force because of $l=0,\chi=0$.
For a finite chromaticity, differences emerge between the cases with and without the space-charge force.
Without the space-charge force, instability occurs at \(-\nu_R=\nu_x+l\nu_z+p-\xi/\eta_p\).
The growth rate decreases to 27\%, 16\%, 12\% and 10\% for \(|l|=1,2,3\) and 4
as shown in Fig.~\ref{foscx26}(b).
The growth rate is independent of the sign of \(l\).

With the space-charge force at \(N_p=7.5\times 10^{12}\) (\(|\Delta\nu_{sc}/\nu_z|=27\)),
instability occurs at \(-\nu_R=\nu_l+p-\xi/\eta_p\), 
where \(\nu_l\) is the tune of an isolated mode related to the $l$-th synchro-betatron mode.
The growth rate decreases to 52\%, 49\%, 46\% and 44\% for \(l=1,2,3\) and 4
as shown in Figs.~\ref{rhox_bt}(b) and \ref{rhox_summary}.
The instability does not occur for \(l<0\) due to Landau damping.
The growth of other bunch intensities \(N_p\) is obtained from the reduction factor
seen in Fig.~\ref{rhox_summary}.

The formulae are extended to the multi-bunch instability with \(H\) bunches
stored with equal spacing (\(C/H\)),
\beqa
\lefteqn{(\nu-\nu_{m}) =-\frac{i H N_pr_p \beta_x}{4\pi\gamma}\omega_0 }\label{nuinst2}\\
&& \sum_p   Z(\omega_{m\mu p}) g_{m,L}^*(\kappa_{m\mu p}-\chi) g_{m,R}(\kappa_{m\mu p}-\chi)\nonumber
\eeqa
where \(\mu=0,1,\ldots,H-1\) is the mode number of the coupled-bunch instability~\cite{achao,cbisc}
and \(\omega_{m\mu p}=(\nu_m+\mu+pH)\omega_0\) (\(\kappa_{m\mu p}=\omega_{m\mu p}\sigma_z/(c\beta)\)).

\section{Conclusions}
Understanding the oscillation modes amounts to representing the transformation of the system
by a matrix and solving an eigenvalue problem.
The beam distribution is represented by a vector of length \(40\times41=1640\), and the
transformation of the beam distribution is represented by a \(1640\times1640\) matrix.
Solving this eigenvalue problem yields the oscillation modes of the bunch;
the beam oscillation is expressed as a linear combination of the eigen-oscillation modes,
and the time evolution and the linear response are analyzed.
Solving the eigenvalue problem relies on numerical methods, but once the eigenvalues and
eigenvectors are obtained, their treatment is standard linear algebra.

Oscillation modes of the beam were discussed for cases in which particles undergo free
synchro-betatron oscillations, and for particles correlated by the space-charge forces.
Eigen-oscillation states were found for both cases, considering the modulation due to
chromaticity in the synchro-betatron modes.
Without space-charge forces, eigen-oscillations exist at the betatron oscillation frequency
and its sideband frequencies \(\nu_x+l\nu_z\) arising from periodic azimuthal displacements
(\(l\)) due to synchrotron oscillations,
and are degenerate with respect to the radial direction in synchrotron phase space.

Space-charge forces lift this degeneracy, causing the tune to depend not only on \(l\)
but also on \(J\). While a group of modes with continuously distributed tunes over \(J\)
appears, for \(l\ge 0\), modes with discrete tunes, separated and isolated from
the continuous tune region, emerge.

These tune distributions have a significant impact on the following phenomena in high-intensity proton rings,
depending on the presence or absence of the space-charge force:
\begin{itemize}
\item Decoherence and recoherence (echo) when a bunch is uniformly displaced
as an initial condition,
\item Beam transfer function when forced oscillation is applied to a bunch,
\item Beam instability caused by a resonator impedance with a high quality factor \(Q\gtrsim 10\).
\end{itemize}

The initial state of a kicked bunch contains a mixture of various eigenmodes
when chromaticity is nonzero.
By representing the initial state as a linear combination of individual eigenmodes
and propagating each mode with its own eigenfrequency,
decoherence and recoherence can be discussed.
The linear coefficients are dominated by isolated modes,
and as the chromaticity \(|\chi|\) increases, isolated modes of higher order \(l\) are included more.

When forced oscillation is applied, modes corresponding to that frequency are excited.
Only isolated modes are excited; continuous modes are not excited due to Landau damping.
That is, \(l<0\) modes are not excited. Excitation depends on integer multiples of
the revolution frequency added to the tune of the forced oscillation.
As the chromaticity \(\chi\) (or \(\xi\)) is changed, the beam response does not change when the frequency
is chosen so as to satisfy \(\nu=\nu_l+p-\xi/\eta_p\).
The strength of the response depends on the space-charge force. The response is several times higher
than that without the space-charge force.

Beam instability is discussed based on the space-charge oscillation modes,
namely, we assume that the oscillation modes are determined by the space-charge force, 
but are not affected by the impedance.
The impedance with a low frequency \(\omega_R\sigma_z/(c\beta)\sim 1\) and a high quality factor 
\(Q\gtrsim 10\) being studied in CSNS and J-PARC meets this condition.
Instability growth and the beam-transfer function are expressed by the same formula.
Only isolated modes with \(l\ge 1\) are excited by the impedance, but continuous modes distributed over \(\nu<\nu_x\)
are not excited due to Landau damping.
The growth rate depends on the space-charge force. The growth for $|\Delta_{sc}/\nu_z|\ge 10$ 
is several times faster than that without the space-charge force.

\section*{Acknowledgements}
This work is jointly supported by the  Guangdong Special Support Program (Project No. 2024TQ08A570)  
and Guangdong Basic and Applied Basic Research Foundation (Project No. 2021B1515120021).

\end{document}